\documentclass[twocolumn,showpacs]{revtex4}
\usepackage{amsfonts}
\usepackage{amsmath}

\usepackage{graphicx}

\begin{document}
\draft
\title{Linear and nonlinear low frequency electrodynamics of
the surface superconducting states in an yttrium hexaboride a single
crystal}

\author{ M.I. Tsindlekht, V.M. Genkin, G.I. Leviev, I. Felner, O. Yuli,
 I. Asulin and O. Millo}
\affiliation{The Racah Institute of Physics, The Hebrew University
of Jerusalem, 91904 Jerusalem, Israel}
\author{M.A. Belogolovskii}
\affiliation{Donetsk Physical and Technical Institute, National
Academy of Sciences of Ukraine, 83114 Donetsk, Ukraine}
\author{N.Yu. Shitsevalova }
\affiliation{Institute for Problems of Materials Science, National
Academy of Sciences of Ukraine, 03680 Kiev, Ukraine}

\begin{abstract}
We report the low-frequency and tunneling studies of yttrium
hexaboride single crystal. Ac susceptibility at frequencies 10 -
1500 Hz has been measured in parallel to the crystal surface DC
fields, $H_0$. We found that in the DC field $H_0>H_{c2}$ DC
magnetic moment completely disappears while the ac response
exhibited the presence of superconductivity at the surface.
Increasing of the DC field from $H_{c2}$ revealed the enlarging of
losses with a maximum in the field between $H_{c2}$ and $H_{c3}$.
Losses at the maximum were considerably larger than in the mixed and
in the normal states. The value of the DC field, where loss peak was
observed, depends on the amplitude and frequency of the ac field.
Close to $T_c$ this peak shifts below $H_{c2}$ which showed the
coexistence of surface superconducting states and Abrikosov
vortices. We observed a logarithmic frequency dependence of the
in-phase component of the susceptibility. Such frequency dispersion
of the in-phase component resembles the response of spin-glass
systems, but the out-of-phase component also exhibited frequency
dispersion that is not a known feature of the classic spin-glass
response. Analysis of the experimental data with Kramers-Kronig
relations showed the possible existence of the loss peak at very low
frequencies ($<5$ Hz). We found that the amplitude of the third
harmonic was not a cubic function of the ac amplitude even at
considerably weak ac fields. This does not leave any room for
treating the nonlinear effects on the basis of perturbation theory.
We show that the conception of surface vortices or surface critical
currents could not adequately describe the existing experimental
data. Consideration of a model of slow relaxing nonequilibrium order
parameter permits one to explain the partial shielding and losses of
weak ac field for $H_0 > H_{c2}$.

\end{abstract}

\date{\today}
\maketitle

\section{Introduction}
There is a growing interest in exploring physical properties of
materials involving boron-cluster compounds because of a wide
variety of applications~\cite{SER}. Due to $sp^2$ hybridization of
valence electrons, large coordination number and short covalent
radius, boron atoms prefer to form strong directional bonds with
various elements. A large number of experimental and theoretical
studies are concentrated on the families of compact B$_{12}$
icosahedrons and B$_{6}$ octahedrons with a large diversity of
electrical and magnetic characteristics. The highest critical
temperatures of the transition to the superconducting state in
MB$_{6}$ and MB$_{12}$ compounds were found in YB$_{6}$ with
$T_c\leq8.4$ K and ZrB$_{12}$ with $T_c\approx 6.0$ K~\cite{FSK}.
Both materials have a highly symmetrical crystal structure
(CaB$_6$type for YB$_6$ and UB$_{12}$ type for ZrB$_{12}$) that can
be described as boron cages in which yttrium or zirconium atoms
develop large vibrational amplitudes with an Einstein-like (nearly
dispersionless) lattice mode. In spite of some common features, the
two crystals have a few distinct physical characteristics: (i) while
YB$_{6}$ is a classical type-II superconductor~\cite{SKUN},
ZrB$_{12}$ (at least, for temperatures above 4.5 K) may be regarded
as a textbook example of type-I superconductor~\cite{GLT}; (ii)
while the superconducting properties are enhanced at the ZrB$_{12}$
surface~\cite{GLT}, they are suppressed in a YB$_6$ surface (see our
tunneling data below). Therefore, ZrB$_{12}$ and YB$_{6}$ samples
may serve as model systems for investigating surface-related
superconducting effects.

 Nucleation of a superconducting phase in a thin surface
sheath, when the DC magnetic field, $H_0$, parallel to the sample
surface decreases, was predicted in 1963 by Saint-James and de
Gennes in their seminal work~\cite{PG}. They showed that the
nucleation occurs for $H_0<H_{c3} = 2.39\kappa H_c$, where $H_c$ is
the thermodynamic critical field and $\kappa$ is the
Ginzburg-Landau(GL) parameter. Experimental measurements confirm
this prediction~\cite{STR,PAS,BURG,ROLL,SWR,OST,HOP}, and it was
found that at low frequencies a sample in a surface superconducting
state (SSS) shows ac losses with a peak whose position with respect
to the DC field depends on the ac amplitude. The peak magnitude
exceeds the losses observed either in the normal state
($H_0>H_{c3}$) or in the bulk superconducting state($H_0<H_{c2}$).
It was also predicted that the $H_{c3}/H_{c2}$ ratio, is
\textit{temperature independent}. In contrast, a \textit{decrease}
of this ratio was found in the vicinity of $T_c$ in several
experiments~\cite{OST,HOP}. This behavior was associated with the
distribution of $T_c$ at the surface~\cite{HU}.

In the last few years the SSS has attracted renewed interest from
various directions~\cite{TS2,JUR,GM,SCOL,LEV2,GLT}. Stochastic
resonance phenomena in Nb single-crystal were observed in the SSS
~\cite{TS2}. In~\cite{JUR} it was assumed that at $H_0\leq H_{c3}$
the sample surface consists of many disconnected superconducting
clusters and subsequently the percolation transition takes place at
$H_{c3}^{c}=0.81 H_{c3}$. The paramagnetic Meissner effect is also
related to the SSS ~\cite{GM}. Voltage noise and surface current
fluctuations in Nb in the SSS have been investigated ~\cite{SCOL}.
SSS' were found also in single crystals of ZrB$_{12}$~\cite{LEV2}.
In agreement with the previous data~\cite{ROLL} it was
demonstrated~\cite{LEV2} that the waveform of the surface current in
an ac magnetic field has a non-sinusoidal character. A simple
phenomenological relaxation model provides the good explanation of
the experimental data for DC fields near $H_{c2}$ only~\cite{LEV2}.
The relaxation rate in this model depends on the ac frequency and
decreased with decreasing $\omega$~\cite{LEV2}. Detailed
experimental study of the linear ac response in the SSS of single
crystals Nb and ZrB$_{12}$ was published recently in our
paper~\cite{GLT}. We showed that ac SSS losses in these materials
could be considered in the achieved experimental accuracy as a
linear ones and for several DC fields the real part of the ac
magnetic susceptibility exhibited a logarithmic frequency dependence
as for a spin-glass system.

In spite of the extensive studies, the origin of low frequency
losses in SSS is not clear as yet. The critical state model
developed for the SSS in~\cite{FINK} implies that if the amplitude
of the ac field, $h_0$, is smaller than some critical value the
losses disappear. The authors of Ref.~\cite{ROLL} claimed that the
experiment on Pb-2$\%$In alloy confirms this prediction. On the
other hand, the observed response~\cite{JUR} for an excitation
amplitude of 0.01 Oe that is considerably smaller than used
in~\cite{ROLL} showed losses in SSS in Nb sample at a frequency 10
Hz. Our measurement on Nb and ZrB$_2$ single crystals also have
shown that the out-of-phase part of the ac susceptibility,
$\chi_1^{\prime\prime}$, was finite at low excitation
level~\cite{GLT}. We consider these results as an indication of the
inadequacy the critical state model for description of the ac
response in SSS. If we assume that the reason for this discrepancy
with experimental data, is the small value of the critical surface
current, which is much smaller than the current amplitudes  at the
surface, then we have to expect a decreasing of the losses
approximately as $1/h_0$ when the amplitude of the applied ac field,
$h_0$, increases. On the contrary, $\chi_1^{\prime\prime}$ increases
with $h_0$. The ac investigation of YB$_6$ samples has some
advantage due to actually ideal type II magnetization curves in this
material that permits one to avoid possible difficulties in the
interpretation of the experimental data. The experiment showed that
near the transition temperatures SSS exist also in the fields below
$H_{c2}$. We found that some features of nonlinear response took
place at a very weak ac field with amplitude $\simeq 0.005$ Oe. The
ac response at the third harmonic of the fundamental frequency did
not leave any room for the perturbation theory. It was proposed that
the losses in SSS are due to the slow relaxation of the order
parameter at the surface and could not be ascribed to surface
vortices. We found that for small $h_0$ in quasilinear approximation
the integral equation with power dependent nuclear governed the time
behavior of the magnetization in ac fields. Some features of the ac
response resemble the ones of spin-glass system but one has to note
that SSS present a different system with its unique properties.

\section{Experimental Details}
\subsection{Sample preparation}

The yttrium hexaboride single crystal was grown by the inductive
floating zone method of a powder sintered rod with an optimal
composition YB$_{6.85}$ under 1.2 MPa of argon. According to the Y -
B phase diagram, composition with the Y:B=1:6 ratio has undergoes
peritectic melting~\cite{MAS} and irrespective of the Y/B ratio the
YB$_4$ single crystal with preferential orientation [001] begins to
grow. After enrichment of the melting zone by boron (flux method
modification) the yttrium hexaboride single crystals with the [100]
orientation grow with the composition of YB$_{5.79\pm 0.02}$ (ESD).
The total impurity concentration is less than 0.001 \% in weight and
the obtained lattice parameter is 4.1001(4)~{\AA} in accordance with
published data~\cite{COM}. These single crystals exhibited a sharp
superconducting transition with $T_c\approx 7.15$ K.

\subsection{DC and ac measurements}

The magnetization curves were measured using a commercial SQUID
magnetometer. In-phase and out-of-phase components of the ac
susceptibility at the fundamental frequency, and the response at the
third harmonic were measured using the pick-up coil
method~\cite{SH,ROLL}. A home-made setup was adapted to the SQUID
magnetometer, and the block diagram of the experimental setup was
published in Ref.~\cite{LEV2}. The crystal ($10\times  3\times 1$
mm$^3$) was inserted into one of a balanced pair of coils. The
unbalanced signal and the third harmonic signal as a function of the
external parameters such as temperature, DC magnetic field,
frequency and amplitude of excitation, were measured by a lock-in
amplifier. The experiment was carried out as follows. The sample was
cooled down at zero magnetic field (ZFC). Then the DC magnetic
field, $H_0$, was applied. The amplitudes and the phases at all
frequencies of both signals were measured in a given $H_0$
(including at zero field). The excitation amplitude,$h_0$,
 was $0.0005\div0.5$ Oe. It is assumed that in $H_0=0$,
and at low temperature, the ac susceptibility equals to the DC
susceptibility in the Meissner state with negligible losses. This
permits us to find the absolute values of the in-phase and
out-of-phase components of the ac susceptibility for all applied DC
and ac fields and for all frequencies. Both $H_0$ and $h_0$ were
parallel to the longest sample axis.

\subsection{Tunneling measurements}

Measurements of the tunneling spectra were carried out using a home
made scanning tunneling microscope. The YB$_6$ single crystal was
mounted inside the cryogenic scanning tunneling microscope and then
cooled down to 4.3 K. The $dI/dV$~vs.~$V$ tunneling spectra
(proportional to the local density of states) were acquired using a
conventional lock-in technique, while momentarily disconnecting the
feedback loop.

\section{Experimental results}

\subsection{Tunneling characteristics}

\begin{figure}
     \begin{center}
    \leavevmode
 \includegraphics[width=0.9\linewidth]{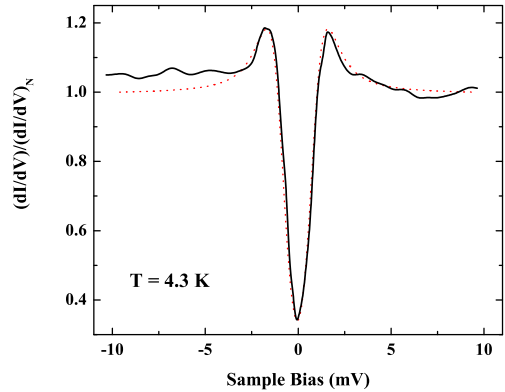}

\caption{Representative tunneling spectrum of YB$_6$ at T 4.3 K
(solid line) together with its fit to the Dynes function (see text)
shown by a dashed line, with fitting parameters $\Delta(T)=1.0$ meV
and $\Gamma =0.10$ meV. The spectra were normalized to the normal
tunneling conductance at 5 meV (well above the superconducting gap)
.}
     \label{f-2B}
     \end{center}
     \end{figure}

Direct information about the energy gap value $\Delta_0=\Delta(T=0)$
at the surface of YB$_6$ was obtained from the tunneling spectra.
The ratio $2\Delta_0/T_c$ is a well known indicator of the
electron-phonon coupling strength~\cite{CARB}. Two previous
tunneling studies of YB$_6$ were performed on a
single-crystal~\cite{SKUN} and on thin films~\cite{SCHN} and the
$\Delta_0$ obtained were: 1.22~\cite{SKUN} and 1.24~\cite{SCHN}
which yields the ratio $2\Delta_0/T_c\approx 4$.  In both cases the
tunneling contacts were connected to underlying layers, and hence,
monitored bulk properties. Therefore those values which signified a
nearly strong coupling are attributed to the bulk characteristics.
It was confirmed, in particular by Lortz \textit{et
al.}~\cite{JUNO}, who measured the deviation function
$D(T)=H_c(T)/H_c(0)-(1-(T/T_c)^2)$ and found that the value of
$2\Delta_0/T_c$ is slightly above 4.0. Our tunneling spectroscopy
results were obtained by STM and therefore better reflect the
density of states at the surface. In contrast to our previous
measurements on ZrB$_{12}$ single crystals that showed very high
spatial homogeneity~\cite{GLT}, the superconductivity in the present
case appeared to be degraded on parts of the YB$_6$ sample surface,
where nearly featureless tunneling spectra were observed. In other
regions, however, reproducible ratios of differential conductances
in superconducting and normal states $(dI/dV)_s/(dI/dV)_n$ showing
very clearly that BCS-like gap structures were acquired, such as
presented in Fig.~\ref{f-2B} (solid line).
The spectra were compared with a temperature-smeared version of the
Dynes formula~\cite{DYN} wich takes into account the effect of
incoherent scattering events by introducing a damping parameter
$\Gamma$ into the conventional BCS expression for a quasiparticle
density of states
 \begin{equation}\label{Eq2B}
N_S(E)=N_N(0)Re[(E-i\Gamma)/\sqrt{(E-i\Gamma)^2-\Delta^2(T)}].
\end{equation}
A very good fit to the experimental data (except for a small
asymmetry in the normal resistance between negative and positive
bias, the origin of which is not yet clear to us) was achieved with
$\Delta(4.3 K) = 1.0$ meV and $\Gamma = 0.10$ meV. Recalling that
the experimental spectrum was acquired at $T = 4.3$ K, which is
about 0.6T$_c$, with the BCS $\Delta(T)$ dependence~\cite{TNK} we
obtain the zero-temperature value $\Delta(T=0) = 1.1$ meV. With
that, we find that $2\Delta_0/T_c\approx 3.59$, very close to the
BCS weak coupling value of 3.53. In contrast to ZrB$_12$, we assume
that in YB$_6$ the electron-phonon strength is suppressed at the
surface to a weak coupling state.

\subsection{DC and ac magnetic characteristics}
\begin{figure}
     \begin{center}
    \leavevmode
 \includegraphics[width=0.9\linewidth]{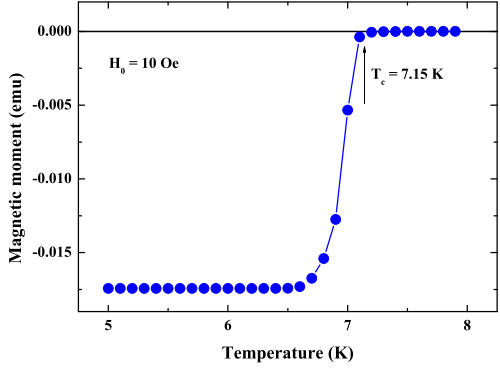}

\caption{(Color online) Temperature dependence of magnetic moment
after ZFC.}
     \label{f-1}
     \end{center}
     \end{figure}

Fig.~\ref{f-1} demonstrates the temperature dependence of the sample
magnetic moment.
In this curve one can see that T$_c\approx 7.15$~K.

\begin{figure}
     \begin{center}
    \leavevmode
 \includegraphics[width=0.9\linewidth]{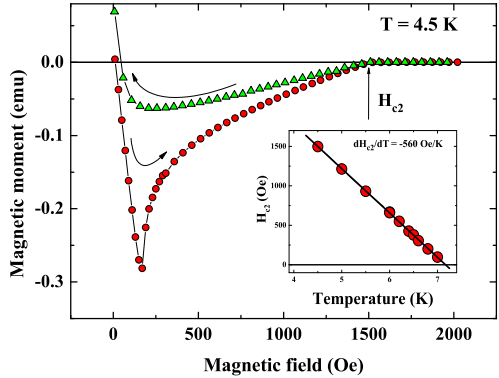}

\caption{(Color online) Magnetization curve at $T=4.5$~K after ZFC.
Inset: temperature dependence of $H_{c2}$.}
     \label{f-2}
     \end{center}
     \end{figure}

From the hysteresis curve measured at 4.5 K, shown in
Fig.~\ref{f-2},
we are able to evaluate H$_c=295$~Oe, H$_{c2}=1500$~Oe and GL
parameter $\kappa_1=H_{c2}/\sqrt{2}H_c =3.58$. Using the relation
$\frac{dM}{dH_0}|_{H_0=H_{c2}}=1/4\pi\beta_A(2\kappa_2^2-1)$ one can
obtain that $\kappa_2= 3.3$, where $\beta_A=1.16$. The temperature
dependence of H$_{c2}$ is shown in the inset of Fig.~\ref{f-2}. The
London penetration depth at T=0, $\lambda_L(0)$, can be estimated by
using $H_{c2}(T)$ near $T_c$, $dH_{c2}/dT\approx -560$~Oe/K,
$1/\lambda_L(0)=\sqrt{\frac{\pi
T_c}{\phi_0\kappa_1^2}|\frac{H_{c2}}{dT}|}$~\cite{ABR}, and
$\lambda_L(0)\approx 1.4\times 10^{-5}$~cm.

\begin{figure}
     \begin{center}
    \leavevmode
       \includegraphics[width=0.9\linewidth]{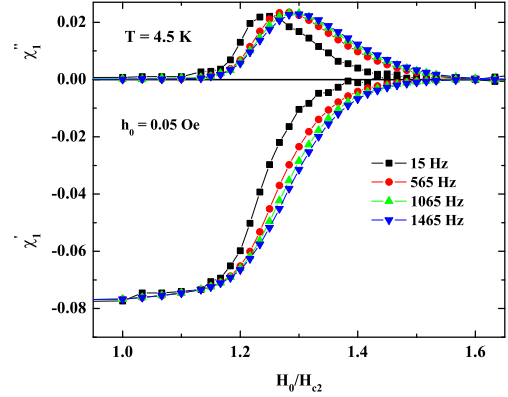}

            \caption{(Color online) Magnetic field dependencies of
    $\chi_1^{'}$ and
    $\chi_1^{''}$
        at $T=4.5$~K at different frequencies $\omega$.}

     \label{f-4a}
     \end{center}
     \end{figure}

\begin{figure}
     \begin{center}
    \leavevmode
    \includegraphics[width=0.9\linewidth]{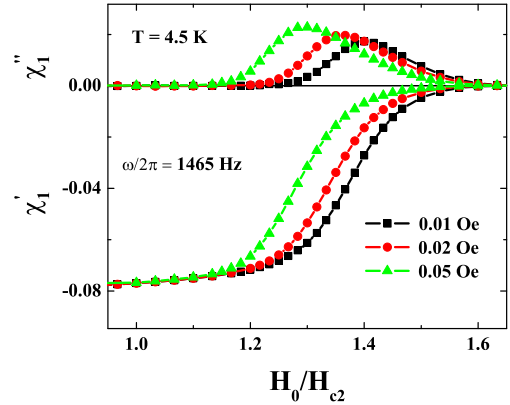}

            \caption{(Color online) Magnetic field dependencies of
    $\chi_1^{'}$ and
    $\chi_1^{''}$
        at $T=4.5$~K at different amplitudes of excitation, $h_0$.}

     \label{f-5G}
     \end{center}
     \end{figure}

Fourier analysis of the magnetization, $M(t)$, under applied ac and
DC fields, $H(t)=H_0+h_0\cos(\omega t)$, yields an expression:
$M(t)=M_0(H_0,h_0)+\sum_n\frac{1}{2}\chi_n(H_0,h_0)h_0\exp(-in\omega
t)$. In this paper we discuss the results for $\chi_1$ and $\chi_3$
susceptibilities. The field dependence of $\chi_1(H_0)$ at $T = 4.5$
K and $h_0 =0.05$ Oe for some frequencies is shown at
Fig.~\ref{f-4a}. One can readily see that the curves shift toward
higher DC fields with frequency.
Decreasing the ac amplitude produces a similar effect. The curves
shift to the higher field when $h_0\longrightarrow 0$ (
Fig.~\ref{f-5G}).
Similar effects were reported for a Pb-2\%In sample in
Ref.~\cite{ROLL}.

\begin{figure}
     \begin{center}
    \leavevmode
 \includegraphics[width=0.9\linewidth]{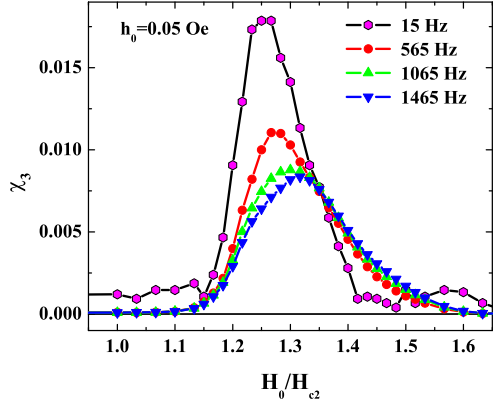}
\caption{(Color online) Third order susceptibility, $\chi_3$, versus
reduced magnetic field, $H_0/H_{c2}$, at different frequencies.}

     \label{f-10G}
     \end{center}
     \end{figure}

\begin{figure}
     \begin{center}
    \leavevmode
 \includegraphics[width=0.9\linewidth]{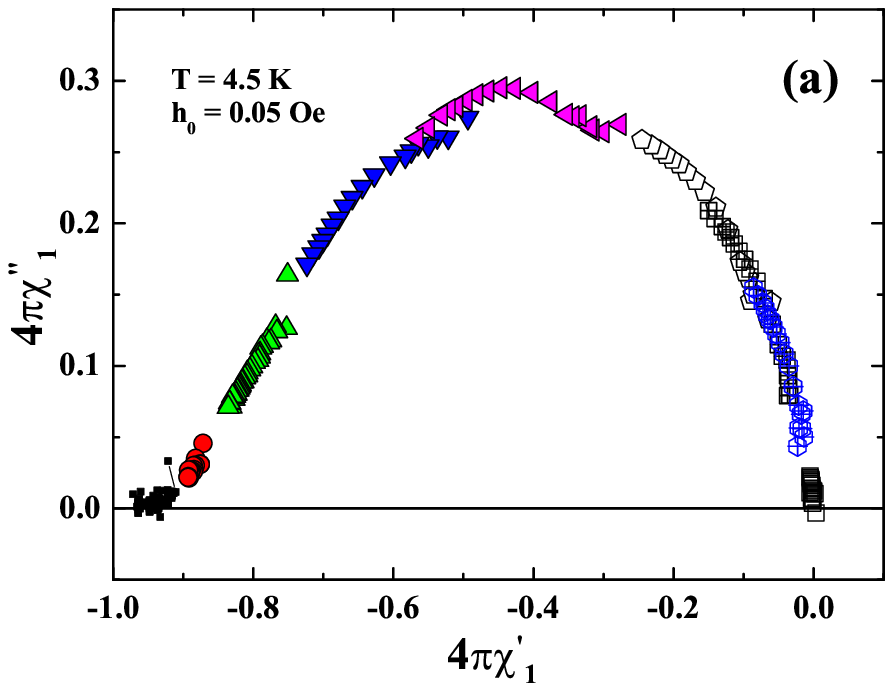}
 \includegraphics[width=0.9\linewidth]{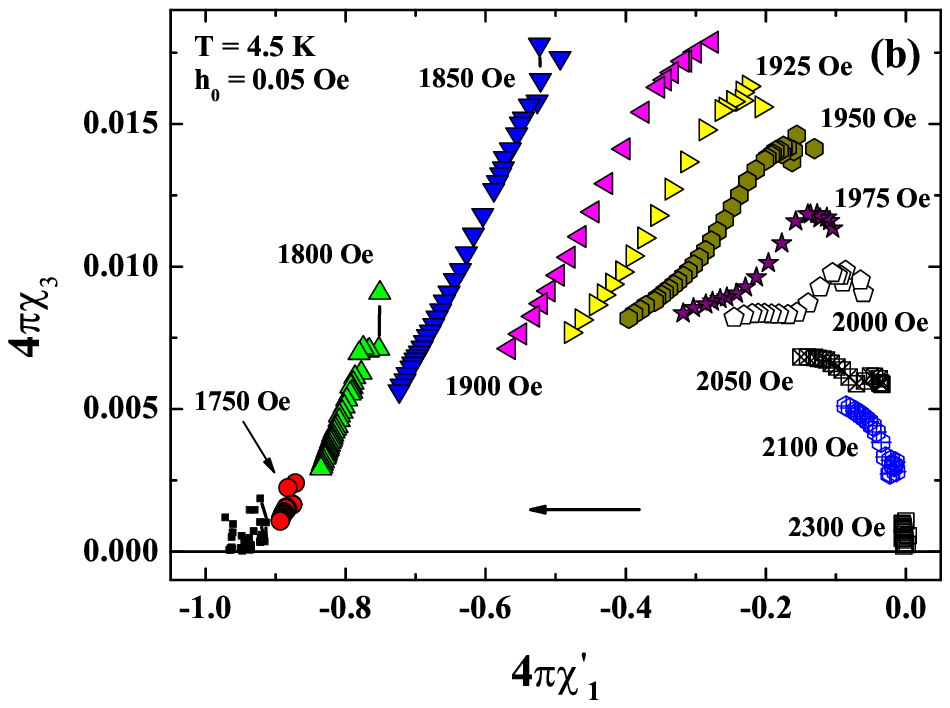}
\caption{(Color online) Panel (a): Cole-Cole plot of the first
harmonic ac susceptibility. Panel (b): $\chi_3$ versus
$\chi_1^{\prime}$. Frequency $\omega$ and DC field $H_0$ are
parameters for these parametric curves. The symbols on the both
panels are the same.}

     \label{f-Xi3G}
     \end{center}
     \end{figure}

The typical magnetic field dependence of the nonlinear response,
$\chi_3$, is presented at Fig.~\ref{f-10G} for $h_0= 0.05$ Oe and
various frequencies.
When the frequency increases the maximum in $\chi_3$  moves toward
larger DC fields as was observed for $\chi_1^{\prime\prime}$ . The
frequency dispersion is illustrated on the Cole-Cole plot,
Fig.~\ref{f-Xi3G}.
One can see $\chi_1^{\prime\prime}$ (panel a) and $\chi_3$ (panel b)
as a function of $\chi_1^{\prime}$ when the frequency increases from
15 to 1465 Hz while the DC field was kept constant. Each
disconnected curve of this figure corresponds to different DC
fields, the values of which are indicated in panel (b). The arrow in
panel (b) shows the direction of increasing frequency along the
curves and shielding, as well as $|\chi_1^{\prime}|$. Below $H_{c2}$
$4\pi\chi_1^{\prime}=-1$ (see Fig. 4). For $H_0>H_{c2}$ both
$\chi_1^{\prime\prime}$ and $\chi_3$ decrease as the frequency
increases while for $H_0$ close to $H_{c3}$ they increase.

\begin{figure}
     \begin{center}
    \leavevmode
 \includegraphics[width=0.9\linewidth]{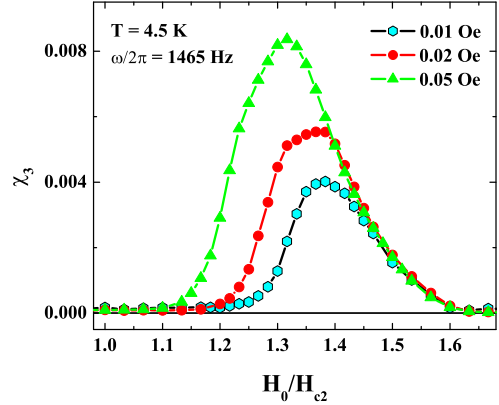}
\caption{(Color online) Third order susceptibility, $\chi_3$, versus
reduced magnetic field, $H_0/H_{c2}$, at different amplitude of
excitation.}

     \label{f-11G}
     \end{center}
     \end{figure}

\begin{figure}
     \begin{center}
    \leavevmode
 \includegraphics[width=0.9\linewidth]{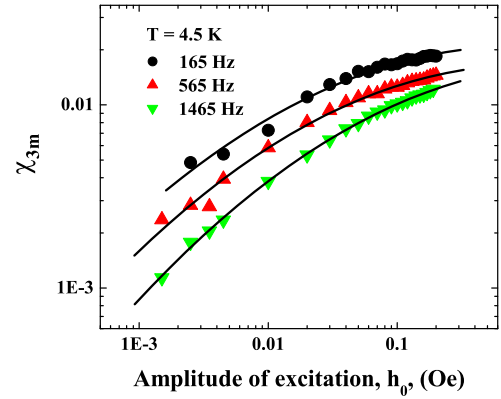}
\caption{(Color online) Amplitude dependence of the third order
susceptibility at maximum, $\chi_{3m}$, at different frequencies
(see text).}

     \label{f-12G}
     \end{center}
     \end{figure}

Fig.~\ref{f-11G} shows the field dependence of  $\chi_3$ at
$\omega/2\pi=1465$ Hz and various amplitudes of excitation, $h_0$.
The third harmonic cannot be adequately described in the frame of
the perturbation theory which predicts that $\chi_3\propto h_0^2$.
For example, at $H_0/H_{c2}=1.3$, $\chi_3$ depends on $h_0$
strongly, while at $H_0/H_{c2}=1.45$, $\chi_3$ is almost constant
(see Fig.~\ref{f-11G}). We can discuss only the dependence of
$\chi_{3m}$ (defined as the maximum value of the $\chi_3(H_0)$ curve
for any given frequency) on the ac amplitude $h_0$. Fig.~\ref{f-12G}
demonstrates that $\chi_{3m}\approx h_0^{0.2}$ in contrast to what
the perturbation theory predictions.

\begin{figure}
     \begin{center}
    \leavevmode
 \includegraphics[width=0.9\linewidth]{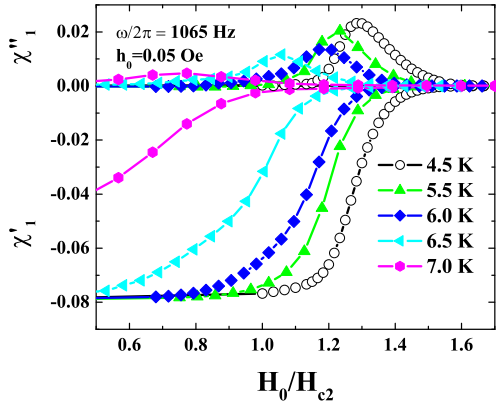}
\caption{(Color online) Field dependence of $\chi_1^{\prime}$ and
$\chi_1^{\prime\prime}$ for various temperatures.}

     \label{f-17G}
     \end{center}
     \end{figure}

\begin{figure}
     \begin{center}
    \leavevmode
 \includegraphics[width=0.9\linewidth]{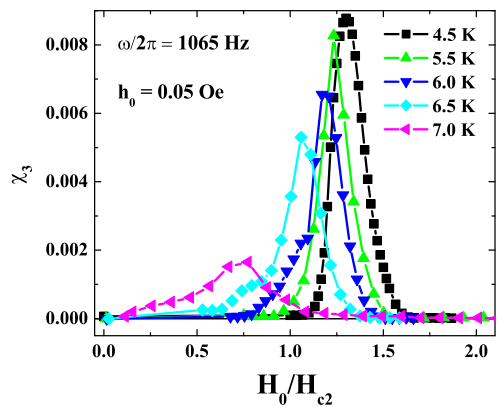}
\caption{(Color online) Field dependence of $\chi_3$
 for different temperatures.}

     \label{f-18G}
     \end{center}
     \end{figure}

Below we consider the experimental results obtained at higher
temperatures. Fig.~\ref{f-17G} demonstrates the field dependence of
$\chi_1$ at frequency $\omega/2\pi= 1065$ Hz and $h_0 =0.05$ Oe at
various temperatures.
The peak in $\chi_1^{\prime\prime}$ shifts toward $H_{c2}$ with
temperature and at 7 K this peak is located already below $H_{c2}$.
One can see in the Fig.~\ref{f-17G} that for $T<7$ K full shielding
($\chi_1^{\prime}=-1/4\pi$) is observed at low $H_0$, whereas at 7 K
only partial shielding is observed at low DC field. Also
Fig.~\ref{f-18G} shows that in the vicinity of $T_c$ $\chi_{3m}$
lies below $H_{c2}$. Because we did not observed any absorption peak
and harmonic signal in the mixed state we consider that SSS are
responsible for the experimental observations at $T=7$ K too.
Existence of the SSS below $H_{c2}$ was predicted by H. Fink in
1965~\cite{HF}.

Increasing the DC field we can reach the field at which $\chi_1$ or
$\chi_3$ becomes zero. This field can be considered as the third
critical magnetic field $H_{c3}$. Both conditions actually give the
same value of $H_{c3}$. The experiment shows that the
$H_{c3}/H_{c2}$ ratio decreases with temperature.

\section{Theoretical model}
Let us consider a superconducting slab of thickness 2$L$ in the
parallel to its surface external DC and ac magnetic fields. Due to
the considerably short relaxation time of the order
parameter~\cite{TNK,TR} one can use the stationary GL equations. We
choose the coordinate system in which the $x$-axis is perpendicular
to the slab surface, the plane $x=0$ is in the center of the slab,
and the external magnetic field is directed along the $z$-axis.
Looking at the dimensionless order parameter in the form
$\Psi(x,y,t)=\phi(x,t)\exp(iky)$ the GL equation can be written as:
\begin{equation}\label{Eq3} 
\ln(T_c/T)\{-\phi+|\phi|^2\phi\}-\frac{d^2\phi}{dx^2}+(a-k)^2\phi=0,Eq2
\end{equation}
\begin{equation}\label{Eq4}
 \frac{d^2a}{dx^2}=\frac{\ln(T_c/T)}{\kappa^2}|\phi|^2(a-k).
\end{equation}
Here $a$ is a $y$-component of the dimensionless vector potential.
The order parameter is normalized with respect to the absolute value
of the order parameter in zero field, the distances with respect to
the coherence length at zero temperature, $\xi_0$,
($x\longrightarrow x/\xi_0,~y\longrightarrow y/\xi_0,~l=L/\xi_0$)
and the vector potential with respect to $\hbar c/2e\xi_0$
($a=A/(\hbar c/2e\xi_0$)). The boundary conditions for calculation
of surface states are $\phi(0,t)=d\phi(\pm l,t)/dx=0$ and
$a(0,t)=0,~da(\pm l,t)/dx=\mathfrak{h}(t)$, where $\mathfrak{h}(t)$
is the dimensionless applied magnetic field.

These nonlinear equations can be solved by numerical methods. We add
the time derivative $\partial\phi/\partial t$ into the right side of
Eq.~(\ref{Eq3}) and seek the stationary solutions of the
Eqs.~(\ref{Eq3}, \ref{Eq4}). Replacing the space derivatives by
finite differences on the grid with step $dx=l/N$ Eq.~(\ref{Eq3})
transforms into $N$ first order differential equations. The solution
of the obtained linear algebraic system can be found by regular
method. The grid with N=1000 points was used. In the surface state
the order parameter differs from zero only near the surface, at a
scale of several coherence lengths, $\xi(T)$. Actually, the choice
$L=5\xi(T)\equiv D$ provides good accuracy for calculating $\phi$.
The real dimensions of the investigated samples, $L$, considerably
exceed this scale by 3-5 orders of magnitude. Parameter $k$ is not a
gauge invariant quantity and we choose it using conditions $a=0$ at
$x=0$. In SSS the magnetic field in the bulk is constant. So we can
obtain $k$ for a thick slab with $L\gg 5\xi(T)$ from the solution of
the problem for a thin slab with $D\geq5\xi(T)$ by gauge
transformation
\begin{equation}\label{Eq5}
k=k_{s}+\mathfrak{h}_{zs}\times(l-d)
\end{equation}
and vector potential in the surface layer
\begin{equation}\label{Eq6}
a(l-d+x)=a_{s}(x)+\mathfrak{h}_{zs}\times(l-d).
\end{equation}
Here $d\equiv D/\xi_0$, index $s$ corresponds to the problem for a
thin slab, and $\mathfrak{h}_{zs}$ is the $z$-component of magnetic
field in the center of the thin slab. This note is important for
numerical calculations.

\section{Discussion}

\begin{figure}
     \begin{center}
    \leavevmode
 \includegraphics[width=0.9\linewidth]{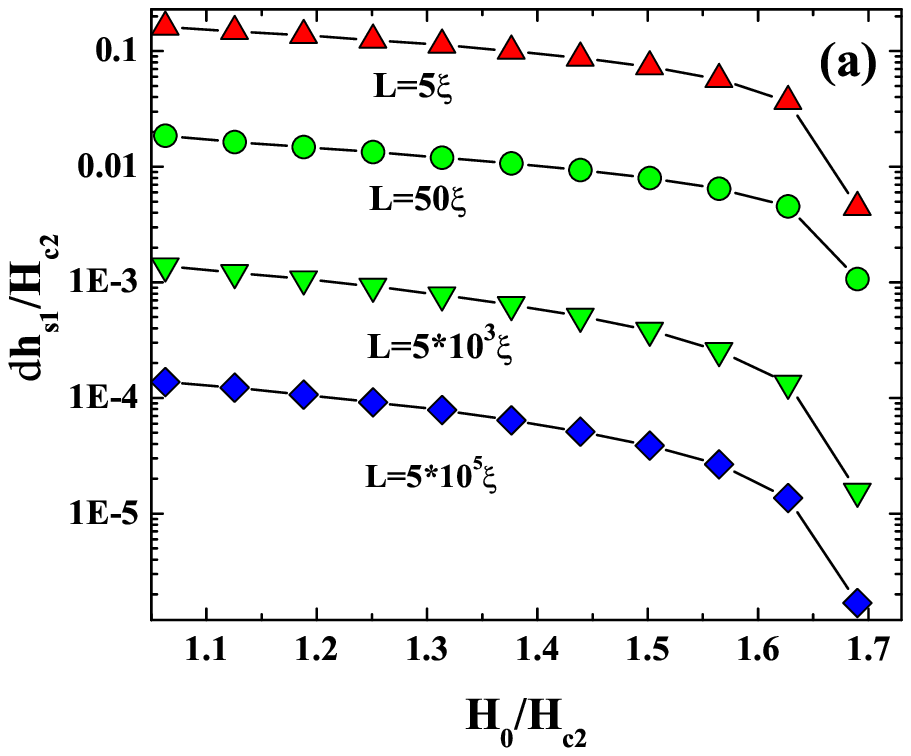}
\includegraphics[width=0.9\linewidth]{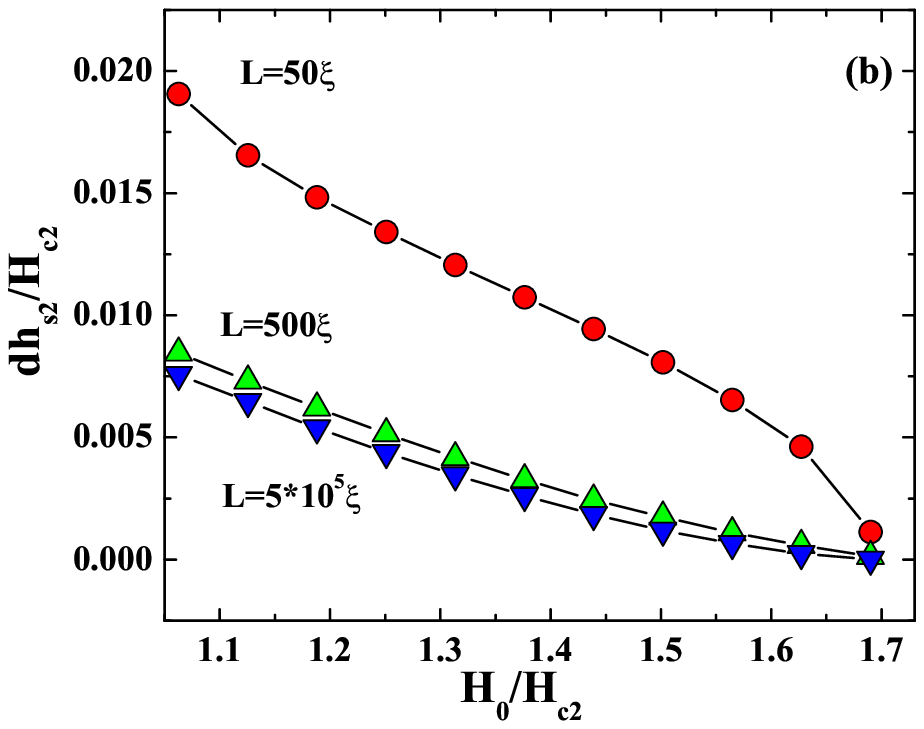}

\caption{(Color online) Field dependence of the surface critical
magnetic field (a) - $dh_{s1}(H_0/H_{c2})$ and (b) -
$dh_{s2}(H_0/H_{c2})$ for different slab thickness, $L$, at
$T/T_c=0.9$ (see text).}
     \label{f-1G}
     \end{center}
     \end{figure}

\begin{figure}
     \begin{center}
    \leavevmode
 \includegraphics[width=0.9\linewidth]{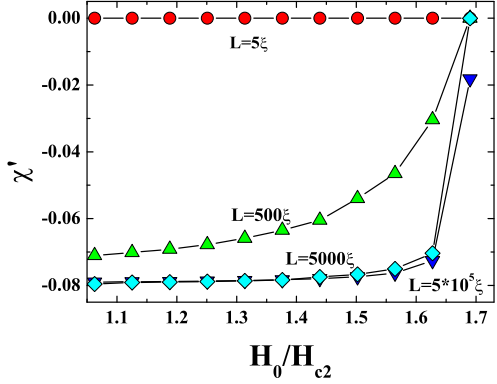}

\caption{(Color online) Field dependence of $\chi'$ for different
slab thickness, $L$, at $T/T_c=0.9$ (see text).}
     \label{f-2G}
     \end{center}
     \end{figure}

It is well known (see, for example,~\cite{LEV2}) that for a given
external magnetic field there is a whole band of $k$  for which
surface solutions exist. These solutions describe the nonequilibrium
states and only one solution corresponds to the equilibrium state,
for which the magnetic field inside the bulk equals its external
value and the total surface current, $J_s$, equals zero. Parameter
$k$ is an integral constant of the nonstationary GL equations. That
is, $k$ is time independent, in contrast to $\phi$, in the frame of
the GL model. The relaxation time of the order parameter $\phi$ is
considerably shorter than any ac period in our experiment. So when
the external magnetic field is changing during the ac cycle, one may
expect that $\phi$ follows the instantaneous value of the magnetic
field and $k$ remains approximately constant. Let assume that
starting from an equilibrium state in some DC field, $H_0$, we
increase the external magnetic field but simultaneously hold $k$
constant. In this case the surface current $J_s$ becomes different
from zero. It is possible to consider two definitions of the surface
critical current $J_{s1}$ and $J_{s2}$~\cite{FINK,PARK}. The first
definition of such a critical current is $J_{s1} =(c/4\pi) dh_{s1}$,
where $dh_{s1}=H_1-H_0$ and $H_1$ is the field for which the energy
of the surface superconducting state equals the energy of the normal
state~\cite{FINK}. The second definition is $J_{s2}=(c/4\pi)
dh_{s2}$, where $dh_{s2}=H_2-H_0$ and $H_2$ is the field for which
SSS disappears. The quantities $dh_{s1}$ and $dh_{s2}$ have
different values and different dependencies on the thickness of the
sample, $L$. While $dh_{s1}$ dramatically depends on $L$, $dh_{s2}$
for $L>1000\xi$ actually does not. The value of $dh_{s2}$ is
considerably larger than $dh_{s1}$ for large $L$. This difference is
due to the large contribution of the magnetic field to the system
energy, if the magnetic field in the bulk differs from the external
field. These features are shown in Figs.~\ref{f-1G}a,~\ref{f-1G}b,
where $dh_{s1}$ and $dh_{s2}$ are presented as a function of the DC
magnetic field for different $L$'s at $T/T_c =0.9$.
In the reduced variables $dh_{s1}/H_{c2}$ , $dh_{s2}/H_{c2}$,
$H_0/H_{c2}$ the curves form is actually temperature independent.
The assumption of slow relaxing $k$, permits one to understand
qualitatively the effect of complete screening of a weak ac field
with amplitude $h_0\ll H_0$ in SSS. Ac surface current $J_s(k, H)$
is a function of the instantaneous values of the external magnetic
field and $k$. This function can be calculated for a thin slab of
several coherence length thickness and then using the gauge
transformation, Eqs.(\ref{Eq5} and \ref{Eq6}), to get a solution for
a thick slab. As a function of $k_{s}$ and $H$, the $J_s(k_{s},H)$
is a slow function of $H$. For example, at $T=0.9T_c$ numerical
calculation gives $\frac{\partial h_{zs}(k_{s},H)}{\partial
H}=0.88+0.19(H_{zs}/H_{c2}-1)$. Where $H_{zs}$ is magnetic field in
the center of the slab. So for a \textit{thin} slab, an almost
complete penetration of the ac field inside the bulk takes place and
the value of the surface current is very small. For a \textit{thick}
slab $k\neq k_{s}$ and the requirement of constant $k$ during the ac
cycle, implicitly means that $k_{s}$ also changed according to
Eqs.~(\ref{Eq5} and \ref{Eq6}). This leads to considerably large
surface currents and to screening of the ac field. In reality, we
have large dimensionless parameter $L/\xi(T)$ that increases ac
field screening. Fig.~\ref{f-2G} demonstrates the calculated (in the
assumption of constant $k$) $\chi'=\Delta M/dh_{s1}$, as a function
of the DC field when the external field was increased by $dh_{s1}$.
It is evident that for any macroscopically large sample,
$L\geq5000\xi$, the complete screening, $\chi'=-1/4\pi$, should be
obtained for DC fields excluding fields close to $H_{c3}$. However
our experiments (Fig.~\ref{f-4a}) do not confirm this conclusion. We
see that $\chi_1'$ in the field $H_0\approx(H_{c2}+H_{c3})/2$
already differs from $-1/4\pi$. It means that slow relaxation of $k$
takes place which leads to the losses and incomplete screening.

\begin{figure}
     \begin{center}
    \leavevmode
 \includegraphics[width=0.9\linewidth]{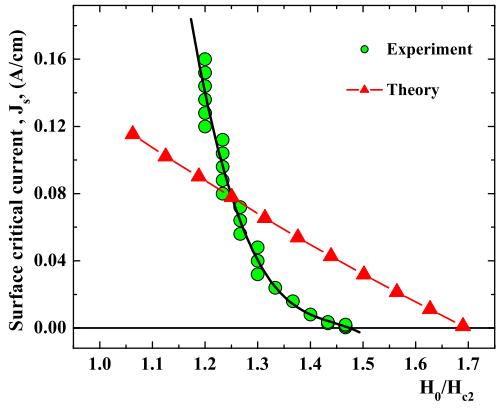}

\caption{(Color online) Surface critical current $J_s$, found with
assumption of Ref.~\cite{BERT}, experimental values and calculated
$J_{s1}$ as a function of reduced DC field $H_0/H_{c2}$ (see text).}
     \label{f-8G}
     \end{center}
     \end{figure}

For a given ac amplitude, $\chi_1^{\prime\prime}$ has a maximum at
some values of the DC field  defined as  $H_m$ (see
Fig.~\ref{f-5G}). $H_m$ was considered in Ref.~\cite{BERT} as the DC
field at which the amplitude of the ac surface current
$J_{0s}=(c/4\pi)h_0$ equals approximately to the critical value
$J_{s1}$. In order to test this in Fig.~\ref{f-8G} we show $J_{0s}$
as a function of $H_0$ and calculate a critical current $J_{s1}$ for
a slab of thickness $L=5\times 10^5\xi$. Theoretical data of the
$J_{s1}$ were arbitrarily normalized in order obtain the
intersection with the experimental curve at $H_0/H_{c2}=1.25$. While
the theoretical dependence of $J_{s1}$ is almost a linear function
of $H_0$, the experimental curve starts from $H_0/H_{c2}=1.45$ and
is a nonlinear function of $H_0$.
One can conclude that losses observed in our experiment are not
connected to the condition $h_0\approx dh_{s1}$ for $H_0>H_{c2}$.

\begin{figure}
     \begin{center}
    \leavevmode
 \includegraphics[width=0.9\linewidth]{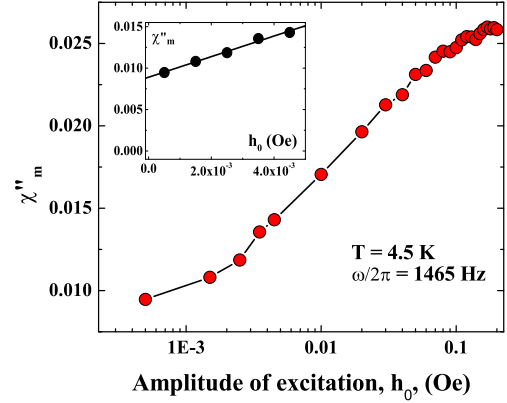}
\caption{(Color online) Out-of-phase susceptibility at maximum,
$\chi_m^{\prime\prime}$, as function of an excitation
amplitude,$h_0$. Inset shows the expanded view for weak $h_0$ in
linear scale.}

     \label{f-9G}
     \end{center}
     \end{figure}

The maximal losses, $\chi_m^{\prime\prime}$, at $H_m$, as a function
of $h_0$ is shown at Fig.~\ref{f-9G}. Inset to Fig.~\ref{f-9G} shows
that in the limit $h_0\rightarrow 0$ the losses do not disappear. In
a linear system $\chi_1^{\prime\prime}$ should be amplitude
independent. While our experiments show a linear dependence
$\chi_m^{\prime\prime}$ on the ac amplitude (Fig.~\ref{f-9G}). It
does not permit us to consider the response as a linear one even at
very low amplitudes of excitation. Therefore more experimental
measurements at low ac fields are needed.

\begin{figure}
     \begin{center}
    \leavevmode
 \includegraphics[width=0.9\linewidth]{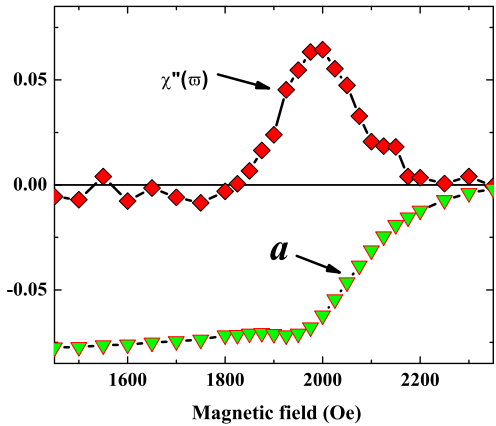}
\caption{(Color online) Field dependencies of
$\chi^{\prime\prime}(\varpi)$ and parameter $a$ of Eq.~(\ref{Eq16a})
at $T=4.5$~K and $h_0=0.02$~Oe (see text).}

     \label{f-13G}
     \end{center}
     \end{figure}

\begin{figure}
     \begin{center}
    \leavevmode
 \includegraphics[width=0.9\linewidth]{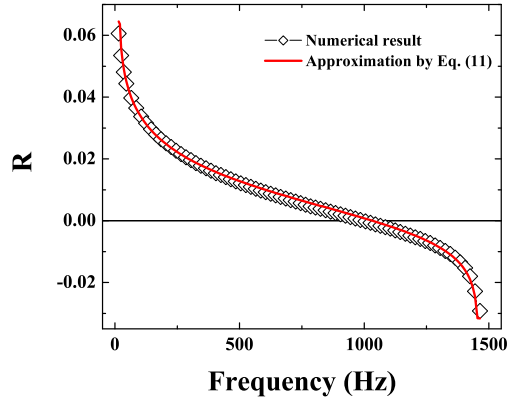}
\caption{(Color online) Frequency dependence of $R$, numerical
calculation and approximation by Eq.~(\ref{Eq17}) (see text).}

     \label{f-16G}
     \end{center}
     \end{figure}

In general, the magnetic moment can be presented by following
expression:
\begin{equation}\label{Eq15}
M(t)=\int_{-\infty}^{t}K(t-t^{\prime},h(t^{\prime}))h(t^{\prime})dt^{\prime}.
\end{equation}
 For $h_0 \simeq 0.02$ Oe the
susceptibilities at higher harmonics are small and we can rewrite
Eq.~(\ref{Eq15}) as

\begin{equation}\label{Eq16}
M(t)=\int_{-\infty}^{t}K(t-t^{\prime},h_0)h(t^{\prime})dt^{\prime}
\end{equation}
considering only the response at the fundamental frequency. Under
this approximation, the response at fundamental frequency, matches
the Kramers-Kronig relations (KKR):

\begin{equation}\label{Eq9}
\chi_1^{\prime}=\chi_{\infty}+\int_0^{\infty}\frac{2\zeta\chi_1^{\prime\prime}(\zeta)}
{\pi(\zeta^2-\omega^2)}d\zeta
\end{equation}
 and then
\begin{equation}\label{Eq10}
\begin{array}{c}
I(\omega)\equiv\chi_1^{\prime}(\omega)-\int_{\omega_0}^{\omega_m}\frac{2\zeta\chi_1^{\prime\prime}
(\zeta)}
{\pi(\zeta^2-\omega^2)}d\zeta =\\
\chi^{\prime\prime}(\varpi)\int_0^{\omega_0}\frac{2\zeta
d\zeta}{\pi(\zeta^2-\omega^2)} +\chi_{\infty}+ \\
\sum_n
\int_{\omega_m}^{\infty}\frac{2\omega^{2n}\chi_1^{\prime\prime}(\zeta)}{\pi\zeta^{2n+1}}d\zeta,
\end{array}
\end{equation}
where $\omega_0$ and $\omega_m$ are the minimal and maximal
available frequencies in our experiment, respectively, and
$0<\varpi<\omega_0$. $I(\omega)$ can be calculated from the
available experimental data and be presented in the form
\begin{equation}\label{Eq16a}
I(\omega)=a+b\ln|1-\omega_0^2/\omega^2|+\sum_{n=1}^{n_{max}}c_n\omega^{2n}.
\end{equation}
With $c_n>0$ we obtain $\chi_1^{\prime\prime}(\varpi)=\pi b$.
Coefficients $a,~ b$ and $c_n$ could be found by least square fit.
For $\omega^2/\omega_m^2<<1$ it is sufficient to take into account
only a few terms in Eq.~(\ref{Eq16a}). Results of this approach are
presented at Fig.~\ref{f-13G}, where $\chi_1^{\prime\prime}(\varpi)$
and $a$ as a function of the DC field are shown.
The measured data in the frequency range 15-1460 Hz $\chi_1$ at
$T=4.5$~K, was used for the calculation of $I(\omega)$ for
$25<\omega/2\pi <200$ Hz with $\omega_0/2\pi =17.5$~Hz and
$\omega_m/2\pi =1455$ Hz. The approximation of $I(\omega)$ by using
expression~Eq.(\ref{Eq16a}) with $n_{max}=1$ produces
$\chi_1^{\prime\prime}(\varpi)$ curve shown in Fig.~\ref{f-13G}.
Because $c_n\approx \omega^2c_{n-1}/\omega_m$ with
$\omega^2/\omega_m^2 \approx 0.02$ one could expect that expression
(\ref{Eq16}) with $n_{max}=1$ gives the correct result. Taking into
consideration the term $c_2$ gives unphysical result, because $c_2$
is very small and negative. It is due to the scattering of the
experimental data and ignores in Eq.~(\ref{Eq16a}) the dependence of
$\varpi$ on $\omega$. Fig.~\ref{f-13G} shows that the calculated
loss peak is approximately 3 times larger than the measured losses
at $\omega/2\pi>20$ Hz, Fig.~\ref{f-4a}. Qualitatively this behavior
can be explained as follows. Because $\chi_1^{\prime\prime}$
exhibits a weak frequency dispersion we can estimate integral in the
left side of Eq.~(\ref{Eq10}) by
\begin{equation}\label{Eq17}
\begin{array}{c}
R=\int_{\omega_0}^{\omega_m}\frac{2\zeta\chi_1^{\prime\prime}(\zeta)}{\pi(\zeta^2-\omega^2)}d\zeta\\
\cong
\chi_1^{\prime\prime}(\omega_m+\omega_0/2)\ln(\frac{\omega_m^2-\omega^2}{\omega_0^2-\omega^2})/
\pi.
\end{array}
\end{equation}
In Fig.~\ref{f-16G} we showed the correspondence between the
estimated $R$ by Eq.~(\ref{Eq17}) and the result of the numerical
calculation of the integral in Eq.~(\ref{Eq17}) at $H_0=2000$ Oe).
It is important that $R$ has a positive sign for $\omega/2\pi<1000$
Hz and in the left side of Eq.~(\ref{Eq10}) one has the sum of two
negative values. So we should expect a large contribution into the
integral in Eq.~(\ref{Eq9}) from frequencies outside the
($\omega_0,~\omega_m$) region and the presentation of this
contribution in the form of Eq.~(\ref{Eq10}) gives a large value for
the term $b$ in Eq.~(\ref{Eq16a}).

We believe that the observed in SSS losses are the result of the
relaxation $k$ to its equilibrium value. This model can ascribe both
the partial screening and losses for $H_0>H_{c2}$. The other model
assumes that the motion of the of 2D-vortices in the surface
sheath~\cite{KUL} is responsible for the losses~\cite{KAR}. These
vortices with surface density $n_s=H_0\sin(\theta)/\phi_0$ appear if
the applied field has a normal component to the sample surface
$H_n=H_0\sin(\theta)$, due to misalignment, or alternatively if the
surface is not sufficiently smooth. One can estimate the
conductivity of the surface layer $\sigma=\sigma_n
H_{c2}/H_0\sin(\theta)$ where $\sigma_n$ is the conductivity in the
normal state. In our sample, $\sigma_n\approx 10^{17}$~CGS and the
skin depth in the surface layer at frequency $\omega/2\pi=10$ Hz is
considerably larger for any real angle( $\simeq10^{-2}$ rad) to
provide sufficient screening of the ac field by a layer with
thickness $10^{-5} \div 10^{-6}$ cm.

\begin{figure}
     \begin{center}
    \leavevmode
 \includegraphics[width=0.9\linewidth]{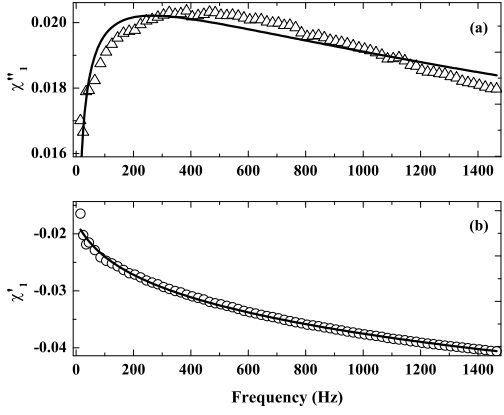}
\caption{Frequency dependence of $\chi_1^{\prime\prime}(\omega)$
(panel (a)) and $\chi_1^{\prime}(\omega)$ (panel (b)) for $H_0=2$
kOe ($H_0=1.25 H_{c2}$ )at 4.5 K. Continuous lines present the fit
to the second order polynomial of $\ln(\omega)$ (Eq.(11)).}

     \label{f-12CG}
     \end{center}
     \end{figure}

\begin{figure}
     \begin{center}
    \leavevmode
 \includegraphics[width=0.9\linewidth]{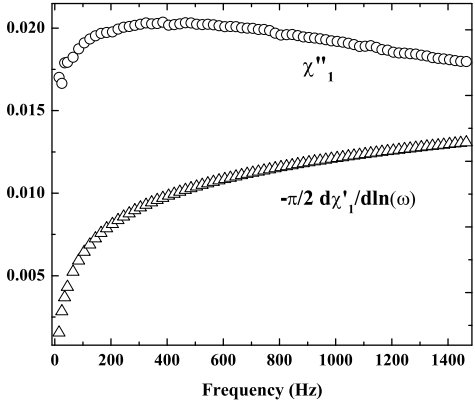}
\caption{(Color online) The test of the "$\pi/2$" rule for the
experimental data at $T=4.5$ K and $H_0=2$ kOe. }

     \label{f-13CG}
     \end{center}
     \end{figure}

The ac response of SSS resembles that of the spin-glass systems.
Real and imaginary parts of $\chi_1$ can be well represented by a
polynomial of $\ln(\omega)$ shown in Fig.~\ref{f-12CG} for $H_0=2$
kOe and T=4.5 K.
In this figure, presentation of $\chi_1\prime$ and
$\chi_1\prime\prime$ by polynomial
   \begin{equation}\label{Eq20a}
a_0+a_1\ln(\omega)+a_2\ln^2(\omega)
\end{equation}
are shown for a considerably wide frequency region
$15<\omega/2\pi<1465$ Hz. For some DC fields the coefficient $a_2$
is small and one can get the spin-glass like $\chi_1^{\prime}$. But
$\chi_1^{\prime\prime}$ also exhibits the frequency dispersion that
is not typical for spin-glass systems. The "$\pi/2$"
rule~\cite{IMRY},
$\chi_1^{\prime\prime}=-\frac{\pi}{2}\frac{d\chi_1^{\prime}(\omega)}{d\ln(\omega)}$,
is not fulfilled in our data, Fig~\ref{f-13CG}.

\begin{figure}
     \begin{center}
    \leavevmode
 \includegraphics[width=0.9\linewidth]{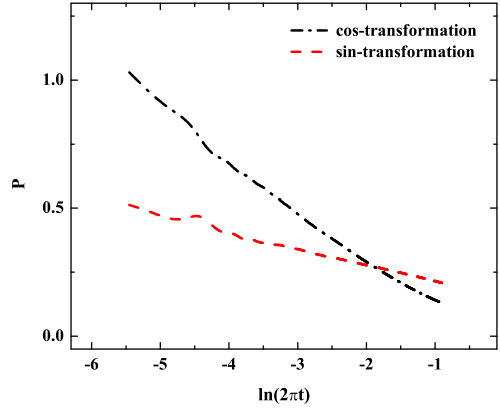}
\caption{The test of the Eq. (\ref{Eq24}) on the experimental data
at $T=4.5$ K and $H_0=2$ kOe.}

     \label{f-14CG}
     \end{center}
     \end{figure}

The simple relaxation models of ac response is applicable only for a
DC field near $H_{c2}$~\cite{LEV2}. If in analogy with a spin-glass
system we assume that the magnetization moment of the sample,
$M(t)$, can be found from the relaxation equation:
\begin{equation}\label{Eq21}
dM/dt=-\nu M-dh/dt,
\end{equation}
with subsequent averaging over the relaxation rates, then
\begin{equation}\label{Eq22}
\chi_1=\int_0^{\infty}\widetilde{P}(\nu)\frac{i\omega}{\nu-i\omega}d\nu,
\end{equation}
where $\widetilde{P}(\nu)$ is the distribution function of the
relaxation rates. Using
$1/(\nu-i\omega)=\int_0^{\infty}\exp(-(\nu-i\omega)t)\text{dt}$ we
transform Eq.(\ref{Eq22}) to
\begin{equation}\label{Eq23}
i\chi_1(\omega)/\omega=\int_0^{\infty}P(\nu)\exp(-\nu t)d\nu,
\end{equation}
where $P(t)=\int_0^{\infty}\widetilde{P}(\nu)\exp(-\nu t)d\nu$. So,
if Eq.(\ref{Eq22}) describes adequately the experimental data with
some $P(\nu)$, then these two integrals should be equal each other
\begin{equation}\label{Eq24}
\begin{array}{c}
P(t)=2\int_0^{\infty}\chi_1^{\prime\prime}(\omega)\cos(\omega
t)d\omega/\pi\omega=\\
-2\int_0^{\infty}\chi_1^{\prime}(\omega)\sin(\omega
t)d\omega/\pi\omega.
\end{array}
\end{equation}
Experimental data are available only for a finite frequency region
15 Hz $<\omega/2\pi<$ 1465 Hz, while integrals in Eq.(\ref{Eq24})
are expanded for all frequencies and we have to extrapolate our data
to the entire frequency axis. This was done assuming that for
$\omega/2\pi< 15$ Hz and $\omega/2\pi>1465$ Hz
$\chi_1{\prime\prime}$ is a power function of frequency $\omega^p$.
As a result, the sin- and cos-Fourier transformations in
Eq.(\ref{Eq24}) give different values for $\widetilde{P}(t)$ as
shown at Fig.~\ref{f-14CG} where the ac response in $H_0=1.25
H_{c2}$ was used.

\begin{figure}
     \begin{center}
    \leavevmode
 \includegraphics[width=0.9\linewidth]{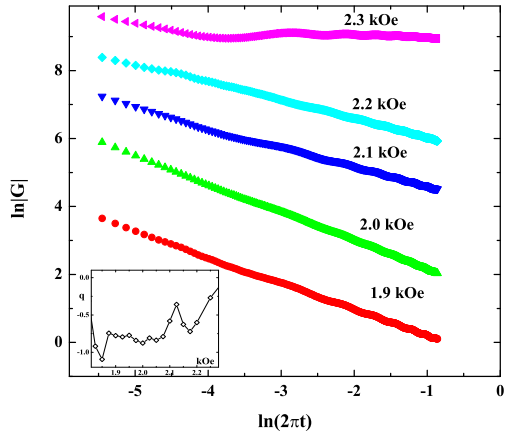}
\caption{Time dependence of nuclear of Eq. (\ref{Eq25}) for several
DC fields near the absorption maximum at $T=4.5$ K.}

     \label{f-15CG}
     \end{center}
     \end{figure}

It is readily seen that the experimental data exclude the
possibility consider the SSS as an analog of a spin-glass system.
Equation (\ref{Eq16}) shows that in the quasilinear approximation
the magnetization of the samples satisfied an integral equation
\begin{equation}\label{Eq25}
\int_{-\infty}^tG(t-t^{\prime},h_0)M(t^{\prime})dt^{\prime}=h(t).
\end{equation}
It is interesting to notice that the nuclear $G(t,h_0)$ can be
extracted by the Fourier transformation of $1/\chi_1(\omega)$.
Performing the same procedure as above, we obtained that sin- and
cos-Fourier transformations in Eq.(\ref{Eq17}) yield different
values for $G(t,h_0)$ which is certainly due to the lack of
experimental data for whole frequency axis. The extrapolation of the
imaginary part of $1/\chi_1(\omega)$ gives more accurate results and
we consider only the $G(t,h_0)$ that is obtained by the sin-Fourier
transformation of $1/\chi_1^{\prime\prime}(\omega)$. Good
approximation of $G(t,h_0)$ provides the expression
$G(t,h_0)=A(t)/t^q$ with slow function $A(t)$ for $t>\pi/1465=t_c$.
For $t<t_c$ function $G(t,h_0)$ is singular, but integral
$\int_0^{t_c}G(t,h_0)\text{dt}$ has a finite value. The parameters
$q$ and $A(t)$, depend on the DC field. For example, in field
$H_0=1.25 H_{c2}$ $q=0.876$ and $A(t)=-\exp(1.285-0.00842\ln^2(2\pi
t))$. In Fig.~\ref{f-15CG} we show $G(t,h_0)$ for some values of the
DC magnetic field and the inset presents $q$ versus $H_0$.
So, the dynamics of SSS is governed by an integral equation with
retardation. This feature distinguishes SSS from other known
systems.

\section{Conclusion}

In this paper we have studied the low frequency linear and nonlinear
dynamics of the SSS of a single crystal of yttrium hexaboride. The
tunneling spectra were studied as well. Tunnel measurements allow us
to make the assumption, that in this single crystal, unlike
ZrB$_{12}$, near the surface the electron-phonon interaction is
suppressed and the situation of weak coupling is realized. We showed
that the surface superconducting states define the peculiarities of
the low frequency response.  In spite of different behavior under
magnetic fields (ZrB$_{12}$ is a type-I superconductor and YB$_6$ is
that of type-II) and different surface properties the two materials
exhibit very similar and universal ac characteristics reflecting the
nature of the SSS. In both cases we observed a nonlinear response
for very weak ac amplitudes (in experiments with YB$_6$ $h_0$ was as
small as 0.005 ~Oe) and the question about the existence of a linear
response is open. An extrapolation of the low-amplitude data did not
reveal a linear regime. Similar to spin-glass systems (where finite
losses at considerably low frequencies exist), the real part of the
susceptibility exhibits a logarithmic frequency dependence at some
DC magnetic field. But the out-of-phase component has a frequency
dispersion. The frequency dispersion in SSS is different from that
of the spin-glass systems. The slow relaxation of the phase of an
order parameter leads to a frequency dispersion of the ac
susceptibility. The analysis of the experimental data by means of
Kramers-Kronig relations allow us to make the assumption of the
presence of the loss peak at frequencies below 5 Hz.

\acknowledgments

This work was supported by the Israeli Ministry of Science (Israel -
Ukraine fund), and by the Klatchky foundation for superconductivity.
We wish to thank E.B. Sonin and I.Ya. Korenblit for many valuable
discussions.


\begin{references}

\bibitem{SER} T.I. Serebryakova and P. D. Neronov, \textit{High-Temperature
Borides} (Cambridge International Science, Cambridge, 2003).

\bibitem{FSK} Z. Fisk, P.H. Schmidt, and L.D. Longinotti,
Mater. Res. Bull. {\bf 11}, 1019 (1976).

\bibitem{SKUN} S. Kunii, T. Kasuya, K. Kadowaki, M. Date, and S.B. Woods.
Solid State Commun., {\bf 52}, 659 (1984).

\bibitem{GLT} M.I. Tsindlekht, G.I. Leviev, V.M. Genkin, I. Felner,
Yu. B. Paderno, and V.B. Filippov, Phys. Rev. B {\bf 73}, 104507
(2006).

\bibitem{PG} D. Saint-James and  P.G. Gennes, Phys. Lett. {\bf 7}, 306
(1963).


\bibitem{STR} M. Strongin, A. Paskin, D. G. Schweitzer,
 O. F. Kammerer, and P. P. Craig, Phys. Rev. Lett.
{\bf 12}, 442 (1964).

\bibitem{PAS} A. Paskin, M. Strongin, P. P. Craig, and D. G.
Schweitzer, Phys. Rev. {\bf 137}, A1816 (1965).

\bibitem{BURG} J. P. Burg, G. Deutscher, E. Guyon, and A.
Martinet, Phys. Rev. {\bf 137}, A853 (1965).

\bibitem{ROLL} R.W. Rollins and J. Silcox, Phys. Rev. 155, 404 (1967).


\bibitem{SWR} H.R. Hart, Jr. and P.S. Swartz, Phys. Rev. {\bf
156}, 403 (1967).

\bibitem {HOP} J.R. Hopkins and D.K. Finnemore, Phys. Rev. B {\bf 9}, 108
(1974).

\bibitem{OST} J.E. Ostenson and D.K. Finnemore, Phys. Rev. Lett.
{\bf 22}, 188 (1969); F. Cruz, M.D. Maloney and M. Cardona, Phys.
Rev. {\bf 187}, 766 (1969).

\bibitem{HU} C.R. Hu, Phys. Rev. {\bf 187}, 574 (1969).

\bibitem{TS2} M.I. Tsindlekht, I. Felner, M. Gitterman, B.Ya. Shapiro,
Phys. Rev. B, {\bf 62} 4073 (2000).

\bibitem{SCOL} J. Scola, A. Pautrat, C. Goupil, L. Mechin, V.
Hardy, and Ch. Simon, Phys. Rev. B {\bf 72}, 012507 (2005). 15

\bibitem{LEV2} G.I. Leviev, V.M. Genkin, M.I. Tsindlekht, I.
Felner, Yu.B. Paderno, V.B. Filippov, Phys. Rev. B{\bf 71}, 064506
(2005).

\bibitem{JUR} J. K\"{o}tzler, L. von Sawilski, and S. Casalbuoni, Phys.
Rev. Lett., {\bf 92}, 067005-1 (2004).

\bibitem{GM} A. K. Geim, S. V. Dubonos, J. G. S. Lok, M. Henin, J.
C. Maan,  Nature {\  bf  396}, 144, (1998).

\bibitem{FINK} H.J. Fink and L.J. Barnes, Phys. Rev. Lett. {\bf 15},
792 (1965); H.J. Fink, Phys. Rev. Lett. {\bf 16}, 447 (1966).

\bibitem{MAS} T.B. Massalski,
\textit{Binary Alloy Phase Diagrams Materials} (ASM International,
Materials Park, OH, 1990).

\bibitem{COM} \textit{Compounds with Boron: System Number 39},
edited by H. Bergman et al., {\bf C11a} of Gmelin Handbook of
Inorganic Chemistry. Sc, Y, La-Lu Rare Earth Elements
(Springer-Verlag, Berlin, 1990).

\bibitem{SH} D. Shoenberg, {\it Magnetic oscillations in metals},
(Cambridge University Press, Cambridge, 1984). 22

\bibitem{SCHN} R. Schneider, J. Geerk and H. Rietschel, Europhys. Lett., {\bf 4}, 845
(1987).

\bibitem{CARB} J.P. Carbotte, Rev. Mod. Phys., {\bf 62}, 1027
(1990).

\bibitem{JUNO} R. Lortz, Y. Wang, U. Tutsch, S. Abe, C. Meingast, P. Popovich,
W. Knafo, N. Shitsevalova, Yu.B. Paderno, and A. Junod, Phys. Rev.
B, {\bf 73}, 024512 (2006).

\bibitem{DYN} C. Dynes, V. Narayanamurti, and J.P.Garno, Phys. Rev. Lett., {\bf 41},1509
(1978).

\bibitem{TNK} M. Tinkham, \textit{Introduction to Superconductivity},
2nd edition (Dover, New York, 2004).

\bibitem{TR} G.I. Leviev, A.V. Rylykov , M.R Trunin, Pis'ma  Zh. Eksp. Teor. Fiz., {\bf 50}, 78 (1989)
(JETP Lett., {\bf 50}, 88, (1989)).


\bibitem{ABR} A.A. Abrikosov, \textit{Fundamentals of the Theory of Metals} (North-
Holland, Amsterdam, 1988).

\bibitem{HF} H. Fink, Phys. Rev. Lett. {\bf 14}, 853 (1965).

\bibitem{BERT} B. Bertman and M. Strongin, Phys. Rev. {\bf 147}, 268
(1966).

\bibitem{KUL} I.O. Kulik, Zh. Eksp. Teor. Fiz. {\bf 52}, 1632 (1967)
[Sov. Phys. JETP {\bf 25}, 1085 (1967)].

\bibitem{KAR} S.Sh. Akhmedov, S.R. Karasik, A.I. Rusinov, Zh. Eksp.
Teor. Fiz., {\bf 56}, 444 (1969).

\bibitem{PARK} J.G. Park, Phys. Rev. Lett. {\bf 15}, 352 (1965).

\bibitem{IMRY} E. Pytte and Y. Imry, Phys.Rev. B {\bf 35}, 1465
(1987).



\end{references}
\end{document}